\documentclass[numberedappendix]{emulateapj}
\slugcomment{Accepted for publication in A\&A}

\shortauthors{Glatt et al.}

\begin{document}
	\title{Ages and Luminosities of Young SMC/LMC Star Clusters and the recent Star Formation History of the Clouds}
	  
	\author{K. Glatt\altaffilmark{1,2}, E. K. Grebel\altaffilmark{1,2} and A. Koch\altaffilmark{3}}
	\altaffiltext{1}{Astronomical Institute, Department of Physics and Astronomy, 
	University of Basel, Venusstrasse 7, CH-4102 Binningen, Switzerland}
	\altaffiltext{2}{Astronomisches Rechen-Institut, Zentrum f\"ur Astronomie der
	Universit\"at Heidelberg, M\"onchhofstr.\ 12--14, D-69120 Heidelberg, Germany}
  	\altaffiltext{3}{Department of Physics and Astronomy, University of Leicester, University Road,
	Leicester, LE1 7RH, UK}
	%\email{kglatt@uni-heidelberg.de, grebel@uni-heidelberg.de, ak326@astro.le.ac.uk}    

\begin{abstract}
In this paper we discuss the age and spatial distribution of young (age$<$1~Gyr) SMC and LMC clusters using data from 
the Magellanic Cloud Photometric Surveys. Luminosities are calculated for all age-dated clusters.
Ages of 324 and 1193 populous star clusters in the Small and the Large Magellanic Cloud have been determined 
fitting Padova and Geneva isochrone models to their resolved color-magnitude diagrams. The clusters cover an age 
range between 10~Myr and 1~Gyr in each galaxy. For the SMC a constant distance modulus of $(m-M)_0$ = 18.90 and a 
metallicity of Z = 0.004 were adopted. For the LMC, we used a constant distance modulus of $(m-M)_0$ = 18.50 and a 
metallicity of Z = 0.008. For both galaxies, we used a variable color excess to derive the cluster ages.
We find two periods of enhanced cluster formation in both galaxies at 160~Myr and 630~Myr (SMC) and at 125~Myr and 
800~Myr (LMC). We present the spatially resolved recent star formation history of both Clouds based on young
star clusters. The first peak may have been triggered by a close encounter between the SMC and 
the LMC. In both galaxies the youngest clusters reside in the supergiant shells, giant shells, the inter-shell regions, 
and toward regions with a high H$\alpha$ content, suggesting that their formation is related to expansion and shell-shell 
interaction. Most of the clusters are older than the dynamical age of the supergiant shells. No evidence for cluster 
dissolution was found. Computed V band luminosities show a trend for fainter magnitudes with increasing age as well 
as a trend for brighter magnitudes with increasing apparent cluster radii.
\end{abstract}

\keywords{Galaxies: Magellanic Clouds - Galaxies: star clusters - galaxies: stellar content}
%\maketitle

%\titlerunning{Glatt et al.}

\section{Introduction}
\label{youngies_intro}

Due to their proximity the Magellanic Clouds (MCs) offer an excellent opportunity to study their spatially resolved star 
formation (SF) histories. SF can be triggered by several mechanisms such as, e.g., the self-induced gravitational collapse 
of a molecular cloud, tidal shocking, a turbulent interstellar medium, or cloud-cloud collisions \citep[e.g., ][]{McKee07}. 
The MCs and the Milky Way (MW) are interacting with each other. The formation of star clusters younger 
than $\lesssim$1~Gyr in the MCs was probably triggered by interactions of the galaxies with each other and with the MW 
\citep[e.g., ][]{yoshi03}. Star clusters may be produced through strong shock compressions induced by close encounters of 
their host galaxies, which causes enhanced star formation. Conversely, the star formation rate decreases again once the 
galaxies recede from each other. Repeated encounters then lead to episodic cluster formation. In the MCs, a correlation 
between young star clusters and putative close encounters with each other and MW has been suggested by, e.g., 
\citet[][G95]{gir95},~\citet[][PU00]{piet00}, and~\citet[][C06]{Chiosi06}. 

Strong tidal perturbations induced by the encounters could also have triggered the formation of clusters 
\citep*[e.g., ][]{Whitmore99} 
in the MCs. Possible orbits of the Small Magellanic Cloud (SMC), the Large Magellanic Cloud (LMC), and the MW have been modeled 
by several authors \citep[e.g., ][]{Bekki05}. They found that it was difficult to keep the Clouds bound to each other for more 
than 1~Gyr in the past. The LMC and the SMC have been part of a triple system together with the Milky Way since at least 
1~Gyr \citep[e.g., ][Kallivayalil et al. 2006a/b]{Bekki05}. It is possible that the Clouds are not a bound system and that 
they are making their first passage close to the MW. 

Interestingly, the cluster formation histories of the LMC and SMC show large differences. In the LMC, two main epochs 
of cluster formation \citep[e.g., ][]{Bertelli92} have been observed that are separated by an ''age gap'' of about 4-9~Gyr 
\citep[e.g., ][]{holtz99,john99,harzar01}, in which no star clusters have formed. The two epochs of pronounced cluster 
formation occurred $>$9~Gyr ago and $\sim$3-4~Gyr ago. In the LMC, a few globular clusters are found that are as old as 
the oldest Galactic globulars \citep{olsen98}. During the past $\sim$4~Gyr, star clusters have been forming continuously 
until the present day. In contrast, the star clusters in the SMC cover a wide range of ages and continued to form over at 
least the last $\sim$10.5~Gyr \citep[e.g., ][]{glatt08a,glatt08b,Parisi08} to the present day. Interestingly, in the SMC 
the cluster formation history appears to have started with a delay since the SMC formed its first and only globular cluster, 
NGC\,121, 2-3~Gyr later than the LMC or the MW \citep[][and references therein]{glatt08a}. The LMC contains about $\sim$4200 
star clusters, while in the SMC $\sim$770 star clusters have formed (and survived). The cluster census is probably still 
incomplete, missing small and faint clusters that are yet to be detected. Ongoing and prospective space-based 
observations may further increase the number of known star clusters. 

The most recent catalog cross-correlating all known objects of the LMC, SMC, and the Magellanic Bridge region was 
published by \citet{bica08b} (B08). However, the cluster sample still is highly incomplete as pointed out by the authors. 
Only for a few clusters in B08's catalog, ages have been determined. For young SMC clusters, \citet{pietrzynski99} (PU99) 
used isochrone fitting on data from the Optical Gravitational Lensing Experiment \citep[OGLE II; ][]{udal98a} to determine 
ages for 93 clusters. \citet{Dieball02} compiled ages for 306 binary cluster candidates in the LMC from a 
variety of literature sources ranging from multiwavelength integrated light studies to isochrone fitting to resolve
clusters. \citet{rafelski05} (RZ05) made use of integrated colors and derived ages for 200 clusters. C06 determined 
ages of 164 associations and 311 star clusters based on data from the OGLE using isochrone fitting. Their sample 
is the largest available catalog with SMC cluster ages. Ages for young LMC clusters have been provided by G95 based on 
integrated colors (624 objects) and by PU00 using isochrone fitting applied to OGLE-II data ($\sim$600 clusters). 

Luminosities have been published for 204 SMC star clusters by RZ05 measuring integrated colors from the Magellanic 
Clouds Photometric Survey (MCPS). \citet{bica96} (B96) published integrated photometry of 624 LMC star clusters
that was based on observations carried out at the 0.61-m telescope at CTIO in Chile and at the 2.15-m CASLEO telescope 
in Argentina.

In the present study we increase the number of age-dated young LMC and SMC star clusters and calculate V-band luminosities.
We aim at improving the understanding of the cluster age distribution of these two irregular galaxies and present spatial 
distribution maps of the star clusters in both galaxies. To achieve this goal, we make use of ground-based data of the 
Magellanic Clouds Photometric Surveys (MCPSs) \citep{zar02,zar04}. In the next Section the observations and data reduction 
are described. In $\S$~\ref{sec:metunddistmod} the distances, reddenings, and metallicities of both the SMC and the LMC are 
given. In $\S$~\ref{sec:agedist} the clusters' age distribution, spatial distribution, and dissolution effects are discussed 
and in $\S$~\ref{sec:youngies_lum} the correlation between the cluster luminosities and age/radius is derived.

\section{Cluster section}
\label{sec:youngies_litdatta}

We made use of three catalogs in order to select clusters and to obtain color-magnitude diagrams (CMDs) of the young SMC 
and LMC star clusters (age $<$1~Gyr). The first catalog was published by B08 \citep[see also ][]{bica95,bica00} and includes 
cluster positions, angular sizes, and object classes for 17,815 objects in the LMC, SMC, and Magellanic Bridge. This catalog 
combines and cross-identifies objects measured on the ESO/SERC R and J Sky Survey Schmidt films and other catalogs 
\citep[e.g., ][]{pietrzynski99}. 

The other two catalogs are from the MCPS presented by \citet{zar02,zar04} containing U, B, V, and I point-source photometry 
of the central 18~deg$^2$ area of the SMC (5,156,057 stars) and of the central 64~deg$^2$ 
area of the LMC (24,107,014 stars). The data were obtained using the Las Campanas 1-m Swope telescope in Chile. The 
V frame was used as reference and only stars detected in both the V and B frame were retained. 
In both catalogs, the limiting magnitude of the photometry is V$\sim$24~mag. The photometry is highly incomplete below 
U=21.5~mag, B=23.5~mag, V=23~mag, and I=22~mag. This means that it is difficult to derive ages of star clusters older 
than $\sim$1~Gyr due to the limited photometric depth of the MCPS, which does not resolve main-sequence turnoff-points 
(MSTOs) of intermediate-age and older clusters.

Each object in B08's catalog is categorized by an object class. Obvious star clusters are indicated by a 'C' and 
emissionless associations by an 'A'. 'CA' and 'AC' indicate intermediate classes between these two types for which a 
classification was not clear. 'NC' refers to small HII regions with embedded star clusters, while 'CN' are clusters 
that show traces of emission. 'DCN' refers to clusters that are decoupled from nebulae. For our study, only objects 
including a 'C'-classification were used (SMC: 765 objects; LMC: 4089 objects), because these objects are well defined 
and not too extended so that cluster stars can be distinguished from the surrounding field stars. Star clusters that 
have very small radii or that are very sparse could not be age-dated (see $\S$~\ref{sec:youngies_ages}). Our resulting 
catalogs contain ages of star clusters between $\sim$10~Myr and 1~Gyr. Clusters whose appearance in the CMDs indicated 
an age older than 1~Gyr were discarded from the sample, because the MSTOs were not resolved. Because we only
age-dated objects including a 'C' in the B08 classification, our samples contain only objects with a minimum age of 
10~Myr, since very young objects are usually classified as associations or nebulae and thus were excluded from our
sample. For many intermediate-age and old SMC and LMC clusters accurate ages have been determined elsewhere in the 
literature mostly based on HST data \citep[e.g., ][]{migh98,olsen98,rich00,pia01,glatt08a,glatt08b}.

\section{Metallicity, Distance Modulus, Depth Extent}
\label{sec:metunddistmod}

\subsection*{\it SMC:} 

The latest low-resolution spectroscopic measurements of SMC star clusters by \citet{Parisi08} showed that the young 
clusters with ages between $\sim$ 0.9 and 2~Gyr have a spread in metallicity of 0.53~dex. The two youngest clusters 
in their sample, Lindsay\,106 and Lindsay\,108, both with ages around 0.9~Gyr, have mean metallicities of [Fe/H]=-0.88 
and -1.05, respectively. Studies based on spectroscopy of six supergiants in the young cluster NGC\,330 
(age$\sim$20-25~Myr; \citet{grebel96}) rendered a metallicity of [Fe/H] = $-0.69 \pm 0.11$~dex 
\citep[e.g., ][]{hill99}, whereas \citet{Gonzalez99} found [Fe/H] = $-0.94\pm0.02$ from high-resolution sprectra of seven
supergiants in this cluster. It has been suggested repeatedly that NGC~330 is more metal-poor than the surrounding
field star population \citep[e.g., ][]{Grebel92,Gonzalez99}. On the other hand, most field star studies agree on 
a value of [Fe/H] = $-0.70 \pm 0.07$~dex \citep[e.g., ][]{Hill97,venn99}. This metallicity corresponds to an isochrone 
model of Z = 0.004, which we used for the age determination. We do not consider a possible metallicity spread here.

A distance modulus of (m-M) = 18.90~mag \citep[$\sim$60~kpc, e.g.,][]{storm04} was assumed for the SMC.
There are no direct determinations of the cluster distances along the line of sight, but it is assumed that the SMC has
a depth extent of up to 20~kpc \citep{Math88,Hatz93,crowl01,lah05,glatt08b}. The clusters analyzed in our study lie in the 
central region of the SMC main body. \citet{Chiosi06} found a variation in the distance modulus of $\sim$0.14~mag assuming 
an elongation of 4~kpc. The resulting error in log(age) is less than 0.05.

\subsection*{\it LMC:} 
\citet{keller06} studied 19 Cepheid variables in the LMC and derived a mean present-day metallicity of [Fe/H] = $-0.34 \pm 
0.03$~dex for this galaxy. This result is in agreement with previous determinations of the metal abundance within
the young population of the LMC \citep[e.g., ][]{luck98,romaniello05} also using Cepheid variables. This metallicity
matches best the isochrone models of Z = 0.008, which were used for the age determination. 

For the LMC, a distance modulus of (m-M) = 18.50~mag \citep[$\sim$50~kpc; e.g.,][]{alves04} was assumed. There are 
indications that the LMC bar may be offset from the plane of the disk by about 2~kpc 
\citep[e.g., ][]{zhao00,subramanian09}. This leads to a variation in the distance modulus of $\sim$0.08~mag. The resulting 
error in log(age) is less than 0.05.

\section{Cluster age distribution}
\label{sec:agedist}

\subsection{Method}
\label{sec:youngies_ages}

\begin{figure}
   \resizebox{\hsize}{!}{\includegraphics{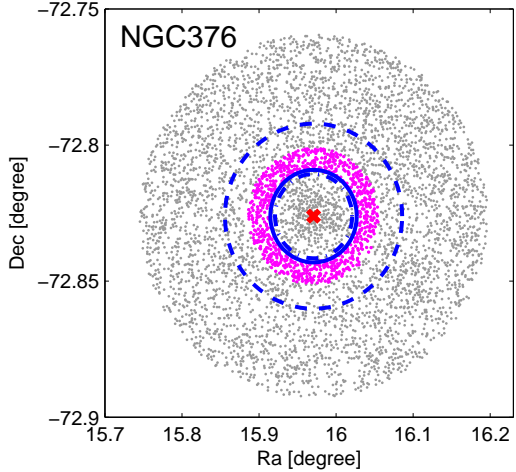}}
      \caption{Spatial distribution of the SMC star cluster NGC~376 (MCPS, V$<$20.5~mag). The red cross denotes the 
      cluster center (B08). The innermost, blue dash-dotted line represents $r_{app}$=0.9~arcmin (B08). The solid blue 
      line shows a circle 1~arcmin away from the cluster center, while the dashed blue line lies 2~arcmin away. Stars 
      lying between these two circles were plotted in the lower panels of Figs.~\ref{fig:cmd376} and~\ref{fig:cmd0899} 
      to visualize the field star populations. The stars plotted as magenta dots are located between $r_{app}+0.1$
      and $r_{app}+0.6$. They correspond to the magenta dots shown in the upper panels of Figs~\ref{fig:cmd376} 
      and~\ref{fig:cmd0899}.}
\label{fig:radii376}
\end{figure}

The cluster age distribution of the LMC and the SMC was obtained by first matching the clusters' central positions 
adopted from B08 with the MCPS \citep{zar02,zar04}. From the apparent major and minor axis adopted by B08 we computed the mean 
apparent diameter

\begin{equation}
D_{\rm app}\,=\,(a + b)/2 
\end{equation}

for each object (see for an example the SMC cluster NGC~376 in 
Fig.~\ref{fig:radii376}). All stars within these diameters are considered cluster members. Four CMDs 
were plotted for each object: Two V vs B-V CMDs and two V vs V-I CMDs showing the cluster and the surrounding field. 

\begin{deluxetable*}{ccccccccl}    
\centering
\tablecolumns{9}
\tablewidth{0pc}
\tablenote{SMC cluster catalog including determined ages and V-band luminosities (full table available online). 
$R_{app}$ denotes the apparent radii (Eq.~1) as adopted from B08. The computed V magnitudes
correspond to the total luminosity of all stars within $R_{app}$ (see Section~\ref{sec:youngies_lum}).}  
\tablehead{            
\colhead{ID} & \colhead{$E_{B-V}$} & \colhead{$R_{app}$} & \colhead{log(age)} & \colhead{$\sigma_t$} & \colhead{Ra} & \colhead{Dec} & \colhead{V} & \colhead{$Cross-ID$} \\
\colhead{} & \colhead{mag} & \colhead{arcmin} & \colhead{yr} &\colhead{} & \colhead{J2000.0} & \colhead{J2000.0} & \colhead{mag} & \colhead{} } 
\hline                       
SMC0017 & 0.02 & 0.25 & 9.00 & 2 & 0:28:31 & -73:00:49 & $16.95 \pm 0.14$ & BS2 \\      
SMC0018 & 0.01 & 0.60 & 8.70 & 1 & 0:30:00 & -73:22:45 & $14.10 \pm 0.13$ & K9, L13\\
SMC0023 & 0.01 & 0.60 & 9.10 & 3 & 0:32:41 & -72:34:53 & $14.77 \pm 0.15$ & L14 \\
SMC0026 & 0.03 & 1.50 & 8.85 & 3 & 0:32:56 & -73:06:58 & $12.51 \pm 0.12$ & NGC\,152, K10, L15, ESO028SC24 \\
\hline                                   
\label{tab:SMC_table}
\end{deluxetable*}

\begin{deluxetable*}{ccccccccl}       
\centering
\tablenote{LMC cluster catalog including determined ages and V-band luminosities (full table available online).
$R_{app}$ denotes the apparent radii (Eq.~1) as adopted from B08. The computed V magnitudes
correspond to the total luminosity of all stars within $R_{app}$ (see Section~\ref{sec:youngies_lum}).}  
\tablehead{            
\colhead{ID} & \colhead{$E_{B-V}$} & \colhead{$R_{app}$} & \colhead{log(age)} & \colhead{$\sigma_t$} & \colhead{Ra} & \colhead{Dec} & \colhead{V} & \colhead{$Cross-ID$} \\
\colhead{} & \colhead{mag} & \colhead{arcmin} & \colhead{yr} &\colhead{} & \colhead{J2000.0} & \colhead{J2000.0} & \colhead{mag} & \colhead{} }
\hline                       
LMC0020 & 0.02 & 0.75 & 8.80 & 1 & 4:37:51 & -69:01:45 & $13.89 \pm 0.14$ & SL8, LW13, KMHK21 \\
LMC0021 & 0.08 & 0.29 & 8.00 & 1 & 4:38:07 & -68:46:39 & $14.29 \pm 0.14$ & NGC\,1649, ESO055SC31, KMHK22 \\
LMC0022 & 0.08 & 0.75 & 8.50 & 1 & 4:38:22 & -68:40:21 & $13.11 \pm 0.14$ & SL14, LW21, KMHK28 \\
LMC0030 & 0.02 & 0.55 & 8.30 & 2 & 4:40:28 & -69:38:57 & $13.73 \pm 0.12$ & SL15, LW23, KMHK29 \\
\hline                                   
\label{tab:LMC_table}
\end{deluxetable*}

We fitted the CMDs with two different isochrone models: Padova isochrones \citep{gir95} and Geneva isochrones 
\citep{Lejeune01}. Both the Padova and the Geneva isochrone grids have an age resolution of log(t)=0.05. The 
Dartmouth isochrone models have a youngest age of 250~Myr \citep{dotter07}, which is too old for many of the star 
clusters in our sample. Therefore, we did not use these models. To fit the isochrones, a constant distance modulus 
and metallicity were used. The best-fit isochrone was then found by using different combinations of reddening and age.
The age and reddening of each cluster are derived by visual inspection. Human judgement is needed in particular 
in order to decide about the inclusion of luminous supergiants in the fit, since the apparent main-sequence 
turnoff of sparse, young clusters is subject to pronounced statistical fluctuations \citep[e.g., ][]{lancon00}. 
The Geneva isochrone models with Z=0.004 deviate from the Padova isochrones for ages younger than log(age)=6.9 by 
about log(age)=$-0.1$. The Geneva models with Z=0.008 deviate by about log(age)=$-0.1$ from the Padova isochrones for 
ages younger than log(age)=8.00. In our catalogs, we list the ages determined using the Padova isochrones. 

\begin{figure}
   \resizebox{\hsize}{!}{\includegraphics{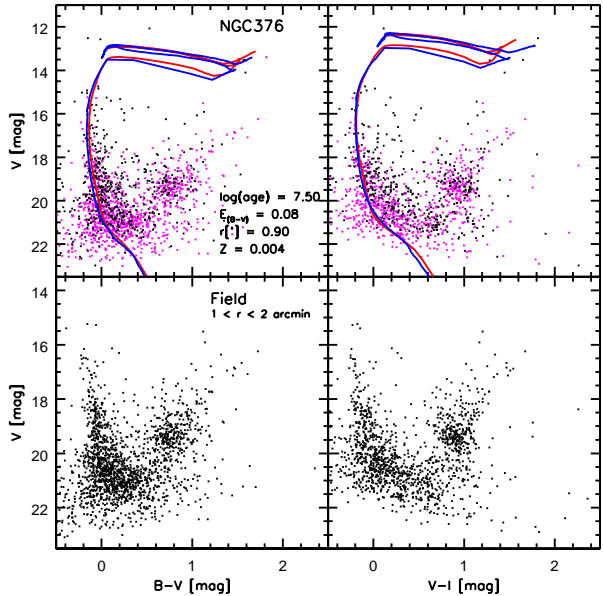}}
      \caption{Color-magnitude diagrams of the SMC star cluster NGC\,376. In the upper panel, V vs B-V and V vs V-I CMDs 
      including stars within a radius of 0.90~arcmin (B08) are shown, and in the lower panel field stars within a 
      concentric annulus between 1 and 2~arcmin are displayed. The blue solid line shows the best-fitting Geneva 
      \citep{Lejeune01} isochrone.The red solid line is the best-fitting Padova \citep{gir95} isochrone. The magenta 
      dots were selected in a 0.50~arcmin concentric annulus that is located 0.1~arcmin away from the cluster apparent 
      radius. They indicate the location of the SMC field stars in the CMDs. The fitted parameters are noted in the 
      Figure.}
         \label{fig:cmd376}
\end{figure}

\begin{figure}
   \resizebox{\hsize}{!}{\includegraphics{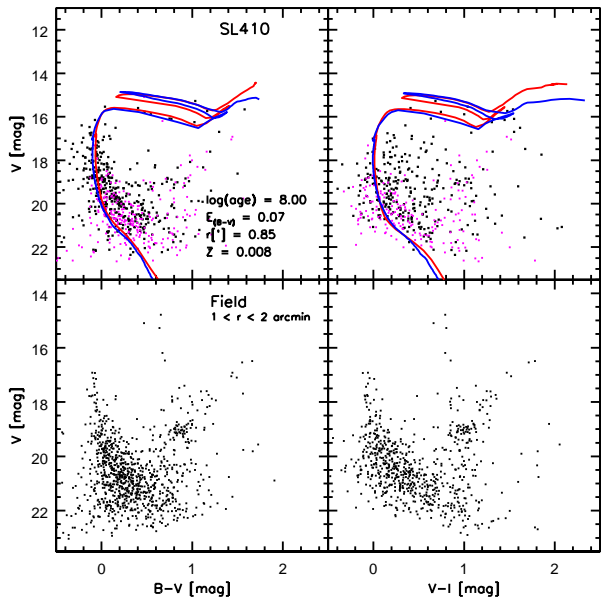}}
      \caption{As in Fig.~\ref{fig:cmd376} but for the LMC cluster SL410.}
         \label{fig:cmd0899}
\end{figure}

Many star clusters in the present samples are located in the main body of the SMC and along the LMC bar, both of which
are highly crowded. Therefore, field star contamination is a severe effect and influences the age determination. 
The field population was sampled within an annulus between 1 and 2~arcmin around the cluster centers outside the 
apparent cluster radii (for examples see lower panels of Figs.~\ref{fig:cmd376} and~\ref{fig:cmd0899}). 
For clusters with larger apparent radii than 1~arcmin, the field population was sampled within an annulus 
between 2 and 3~arcmin. The selected 
stars lying within these annuli were plotted on top of the cluster CMDs to illustrate the location of the SMC field 
stars. The accuracy of the inferred ages depends on the number of cluster member stars and on the density of the 
surrounding field. Many star clusters in the sample could not be age-dated, because they are too sparse or they 
have very small radii, which include too small a number of stars against the SMC field background for a reliable 
isochrone fit. In the densest SMC regions, the cluster membership determination on the basis of photometric 
information is difficult, and therefore the here derived ages are uncertain. 

In Figures~\ref{fig:cmd376} and~\ref{fig:cmd0899} two examples are shown. For the SMC cluster NGC\,376
we derive an age of $\sim$30~Myr and $E_{B-V} = 0.08$. The LMC cluster SL410 is a slightly older star cluster for which 
we determine an age $\sim$110~Myr and $E_{B-V} = 0.07$. In the respective upper panels, stars that were selected 
within the cluster radii adopted from B08 are displayed. The magenta dots represent the field stars selected within a 
concentric 0.5~arcmin wide annulus located 0.1~armin away from the clusters apparent radii.  
The blue solid line shows the best-fitting Geneva isochrone. The red solid line is the best-fitting 
Padova isochrone. In the lower panels, field stars selected within an annulus between 1 and 2~arcmin 
radius around the cluster centers are shown. 

Young cluster ages derived from isochrone fitting are not severely affected by the discreteness of isochrones 
due to the logarithmic age-steps of the applied isochrone models, which produce a fairly dense grid for small ages. 
The interstellar extinction, however, can have a severe impact on the determination of the cluster ages. We compared 
the reddenings found by isochrone fitting to the reddenings adopted from PU99. The standard deviation is of the order
of $\sigma E_{(B-V)}$=0.03. 

In Section~\ref{sec:metunddistmod}, the uncertainty in the present-day metallicity of the 
SMC is mentioned. Consequently, for some clusters, the applied isochrone metallicity of Z=0.004 might not be 
the best choice. However, the attempt to fit the clusters with a different metallicity (e.g., Z=0.008) provided 
fits of lower quality (age difference of $\sim$log(age)=0.2). Therefore, the uncertainties of the age determinations 
are partly a function of interstellar extinction, partly a function of the age itself in the sense that older 
clusters are more difficult to identify, and partly a function of the cluster density. 

An uncertainty that cannot be handled with our data occurs when the clusters are all inside large groups of associations 
having small separations along the line of sight or when they are very sparse.  Sparse objects usually do not
have a well-defined main sequence turn-off. On one hand, supergiants with their short lifetimes (typically $\sim$20~Myr) 
are well-suited
as age tracers if they show a concentration toward the cluster center. We therefore require our "best" isochrones to
reproduce the location to the supergiants in the CMDs and use that as a primary criterion for age determinations. The reddenings,
on the other hand, are based on the perceived locus of the blue envelope of the main sequence. The determination of cluster
membership for individual stars is a difficult matter (based on only photometric information) and further investigation 
with e.g., the Hubble Space Telescope or studies based on spectroscopy are necessary. A clear separation between
clusters and their surroundings becomes particularly difficult when they are embedded in associations of a very similar
age.
 
For the SMC and the LMC, ages of 324 clusters and 1193 clusters, respectively, have been determined in our study. 
The full tables will be made available electronically, but illustrative excerpts are shown in Tables~\ref{tab:SMC_table} 
and~\ref{tab:LMC_table}. Column 1 gives the cluster identification number. The reddening parameter $E_{B-V}$ is listed in column 2 
and the apparent radii in column 3. The derived ages are shown in column 4, while column 5 gives the degree of 
reliability of our age measurements. Class 1 indicates having errors $\Delta \sigma_{log(age)} < 0.3$; class 2
indicates objects having errors  $0.3 \leqslant \sigma_{log(age)} < 0.5$; and class 3 indicates objects having errors
$\sigma_{log(age)} \geqslant 0.5$, adopting the same notation as used by C06. Columns 6 and 7 give the right ascension 
and the declination, respectively. The calculated total V band magnitudes (see $\S$~\ref{sec:youngies_lum}) are listed in 
column 8 and finally, the cross-identifications with other catalogs are given in column 9.

\subsection{Comparison of our age determination with previous studies}
\label{youngies_comparison}

C06 used data from OGLE~II \citep{udal98a} for the 
SMC disk and data obtained at the ESO 2.2~m telescope for the region around NGC\,269 that is located at the border of 
the supershell 37~A to derive ages of 461 SMC clusters and associations. The ages were derived using Padova isochrones 
\citep{gir02}. A distance modulus of (m-M) = 18.9 was assumed and the reddening was derived by main-sequence fitting. 
A mean SMC metallicity of Z = 0.008 was assumed (in the present study we used Z = 0.004) as found spectroscopically 
for young stars by \citet{pagel99}. An upper age limit was set by the limiting magnitude of the available photometry 
indicating that clusters that have an MSTO fainter than V=20~mag in the OGLE~II field could not be age-dated. C06's 
study was restricted to clusters younger than 1~Gyr. They derived ages within an age range of 4~Myr and 1~Gyr for 
clusters covering an area of 2.4~deg$^2$ of the SMC main body. 136 of their clusters are also included in our study. 
 
\begin{figure}[h!]
 \resizebox{\hsize}{!}{\includegraphics{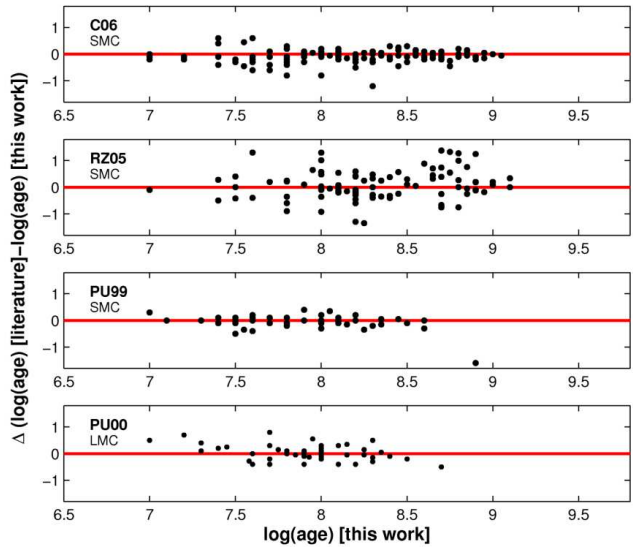}}  
 \caption{Cluster ages derived in this work are compared with the SMC cluster ages by C06 (first panel), RZ05 (second panel), 
 and PU99 (third panel), and with the LMC cluster ages by PU00 (fourth panel) for the clusters in common.}
 \label{fig:ageallvergleich}
\end{figure}

In the first panel of Fig.~\ref{fig:ageallvergleich} cluster ages determined in our work are compared to ages presented  
by C06. The ages derived in our work tend to be $\sim$0.2-0.3 in log(age) older than the ages by C06, an offset we 
anticipated above. The main reason for the difference is probably the different metallicity of the applied isochrone 
models as we tested for a subset of CMDs. For those clusters with the most significant age deviation large age uncertainties 
are stated both in their and our work (e.g. BS271, BS272, B114). The dispersion about the zero line (red solid line) for the 
sample is $\sigma_{log(age)}$ = 0.13.

RZ05 derived ages of 204 SMC star clusters using integrated colors, of which 112 clusters are included in our sample. 
The second panel of Fig.~\ref{fig:ageallvergleich} shows the comparison between our ages and ages adopted by RZ05. 
The largest deviations are found at the ends of the upper and the lower age limits. A few clusters having ages younger 
than 25~Myr and older than 1~Gyr show an age deviation of up to 1~Gyr. Stochastic effects on the number of bright stars, 
uncertainties on the metallicity, dubious cluster membership, and on the adopted stellar models may contribute to the large 
uncertainties in the conclusion of the cluster age. RZ05 used a metallicity of Z = 0.004. Those clusters with the 
largest age differences are mostly very sparse objects or objects to which very young isochrones were fitted based on 
only 2-3 bright stars. Large uncertainties for the oldest clusters are caused by large field contamination and unresolved
main-sequence turnoff points. The dispersion about the zero line (red solid line) for the sample is 
$\sigma_{log(age)}$=0.3. 

PU99 determined ages for 93 SMC star clusters from the OGLE catalog \citep{piet98}, of which 51 are included in our sample. 
They fitted isochrones from the library of \citet{bertelli94} applying a distance modulus of $(m-M)_0$=18.65 \citep{udal98b}
and a metallicity of Z=0.004. In their study, they derived cluster ages within an age range of 10~Myr and 1~Gyr. In the third 
panel of Fig.~\ref{fig:ageallvergleich} we compare ages presented by PU99 to our study. The ages are in very good agreement 
and the dispersion about the zero line (red solid line) is $\sigma_{log(age)}$=0.17. This result may seem surprising
considering the difference of the applied distance moduli of 0.25. However, fitting isochrones to the same CMD using both values 
for the distance moduli leads to age differences for these young clusters of no more than 0.1-0.15 in log(age).

Ages of about 600 LMC star clusters were determined using OGLE II data \citep{udal98a} by PU00. The ages were
derived fitting isochrones of \citet{bertelli94}. A distance modulus of 18.24~mag and a metallicity
of Z = 0.008 were assumed. The reddenings were adopted from \citet{udal1999}. Clusters older than $\sim$1.2~Gyr
could not be reliably age-dated, because the main-sequence turnoff points are located close to the limit of the OGLE~II
photometry (V$\approx$21.5~mag). 49 clusters are in common with our cluster sample. In the fourth panel of 
Fig.~\ref{fig:ageallvergleich} we compare the derived ages found in this study with PU00. The dispersion about the 1:1 
agreement (red solid line) for the sample is $\sigma_{log(age)}$=0.15. Again the good agreement is surprising, 
considering the difference of the applied distance moduli of 0.26. As for the SMC, the age difference for these young clusters
is no larger than 0.1-0.15 in log(age) fitting two isochrones to the same CMD using both distance modulus values.

G95 used integrated colors to determine the S parameter for all 624 LMC objects listed in the catalog of  \citet{bica96}. 
The S parameter is an age indicator based on integrated U-B and B-V colours \citep[see also ][]{els85}. 

Unfortunately, a direct comparison of our data is not possible because the cluster ages of G95 are not tabulated. 
\citet{grebel99} fitted isochrones to data from earlier versions of the MCPS \citep{zar97} and OGLE \citep{udal98a} and 
transformed G95's S parameter into ages. The authors found a good overall agreement in the age distribution.

All in all, the scatter of $\sim$0.2 in age derived by other groups and in this work  is in good agreement with the typical 
age uncertainties derived in either study.

\subsection{Age Distribution}
\label{sec:youngies_agedistribution}

\begin{figure}[h!]
 \resizebox{\hsize}{!}{\includegraphics{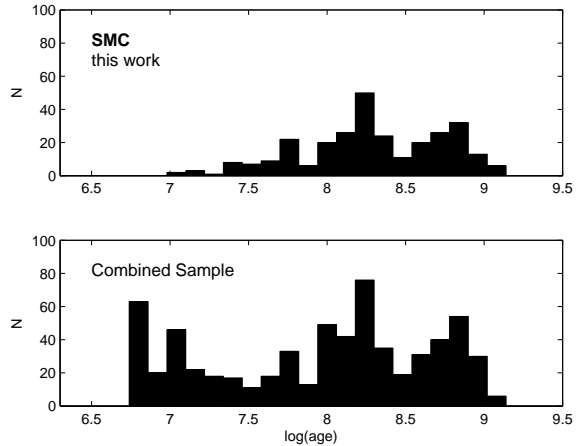}}  
 \caption{Cluster age distribution of the SMC. The age distribution derived in this study (upper panel)
 and the combined samples (lower panel) of this study and C06 are shown (in total 821 clusters). Only the most reliable ages 
 were considered for this plot (classes 1 and 2 in Table~\ref{tab:SMC_table}).}
 \label{fig:agedist_smc}
\end{figure}

\begin{figure}[h!]
 \resizebox{\hsize}{!}{\includegraphics{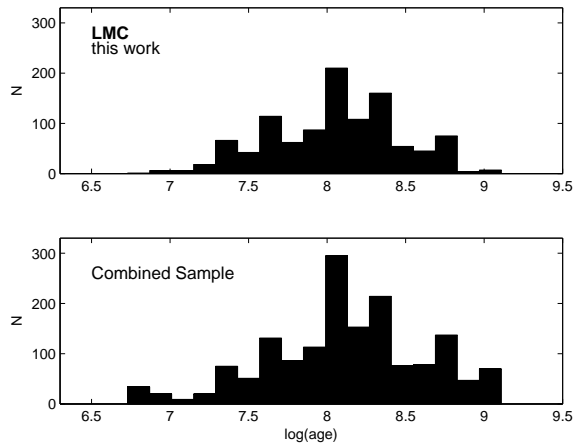}}  
 \caption{Cluster age distribution of the LMC. The age distribution derived in this study (upper panel)
 and the combined samples (lower panel) of this study and PU00 are shown (in total 1745 clusters). Only the most reliable 
 ages were considered for this plot (classes 1 and 2 in Table~\ref{tab:LMC_table}).}
 \label{fig:agedist_lmc}
\end{figure}

Uncertainties aside, the age distribution of star clusters is influenced by two effects \citep{boutloukos03}. The first is 
cluster fading that occurs through stellar evolution when the clusters get fainter with time. Therefore, the number of 
observed star clusters appears to decrease with increasing age for a given magnitude limit since sparse, faint clusters 
become increasingly harder to detect. A second effect acting upon gas-free clusters is caused by the combined 
effects of 2-body relaxation and external tidal stripping: this is the so-called secular evolution \citep{BK03}. 
A third effect altering the initial age distribution at the lower end is infant mortality caused by gas expulsion. 
In the MCs the phase of infant mortality lasts up to $\approx$40~Myr \citep{deGrijs09}.  

The cluster age distributions of the SMC and the LMC are shown in Fig.~\ref{fig:agedist_smc} and~\ref{fig:agedist_lmc}. 
In the upper panels of these figures only cluster ages determined in our study and in the lower panels the combination of 
this study with C06 (SMC) and PU00 (LMC) samples are displayed. For the clusters in common with our work ages derived in 
this study were used. Because we age-dated only objects that included a 'C' in the B08 classification, our samples contain 
only a few objects younger than $\sim$20~Myr, since very young objects are usually classified as associations or nebulae and 
thus were excluded from our sample. 

\begin{figure*}
 \resizebox{\hsize}{!}{\includegraphics{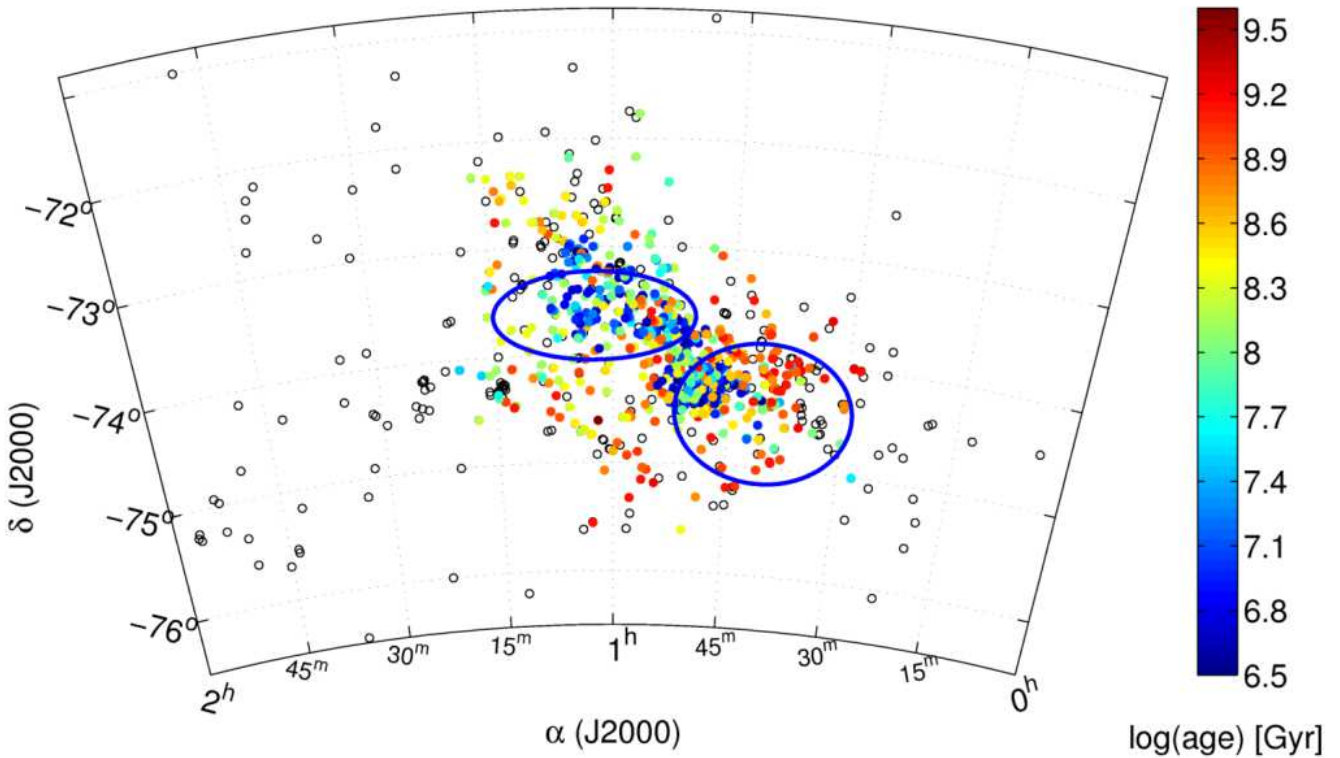}}  
 \caption{Spatial Distribution of the combined sample of this work and C06 of the young SMC star clusters. The two blue 
 ellipses indicate the location of the HI super-shells 304~A (middle) and 37~A (right) \citep{stanimirovic1999}. 
 The open circles indicate all objects from B08 including a 'C'-classification (in total 765 objects). North is up, west 
 to the right.}
 \label{fig:smcageall}
\end{figure*}

\begin{figure*}
 \resizebox{\hsize}{!}{\includegraphics{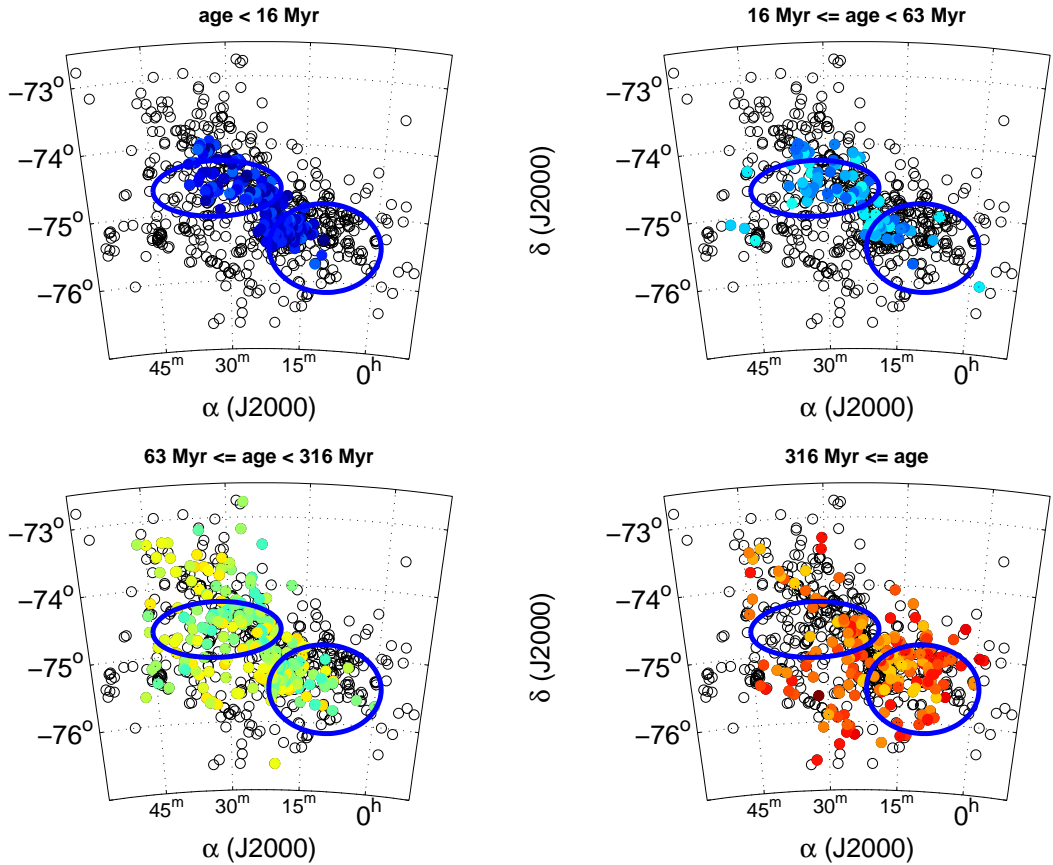}}  
 \caption{Spatial distribution of SMC clusters younger than 1~Gyr in different age bins.}
 \label{fig:smcageslides}
\end{figure*}

\begin{figure*}
 \resizebox{\hsize}{!}{\includegraphics{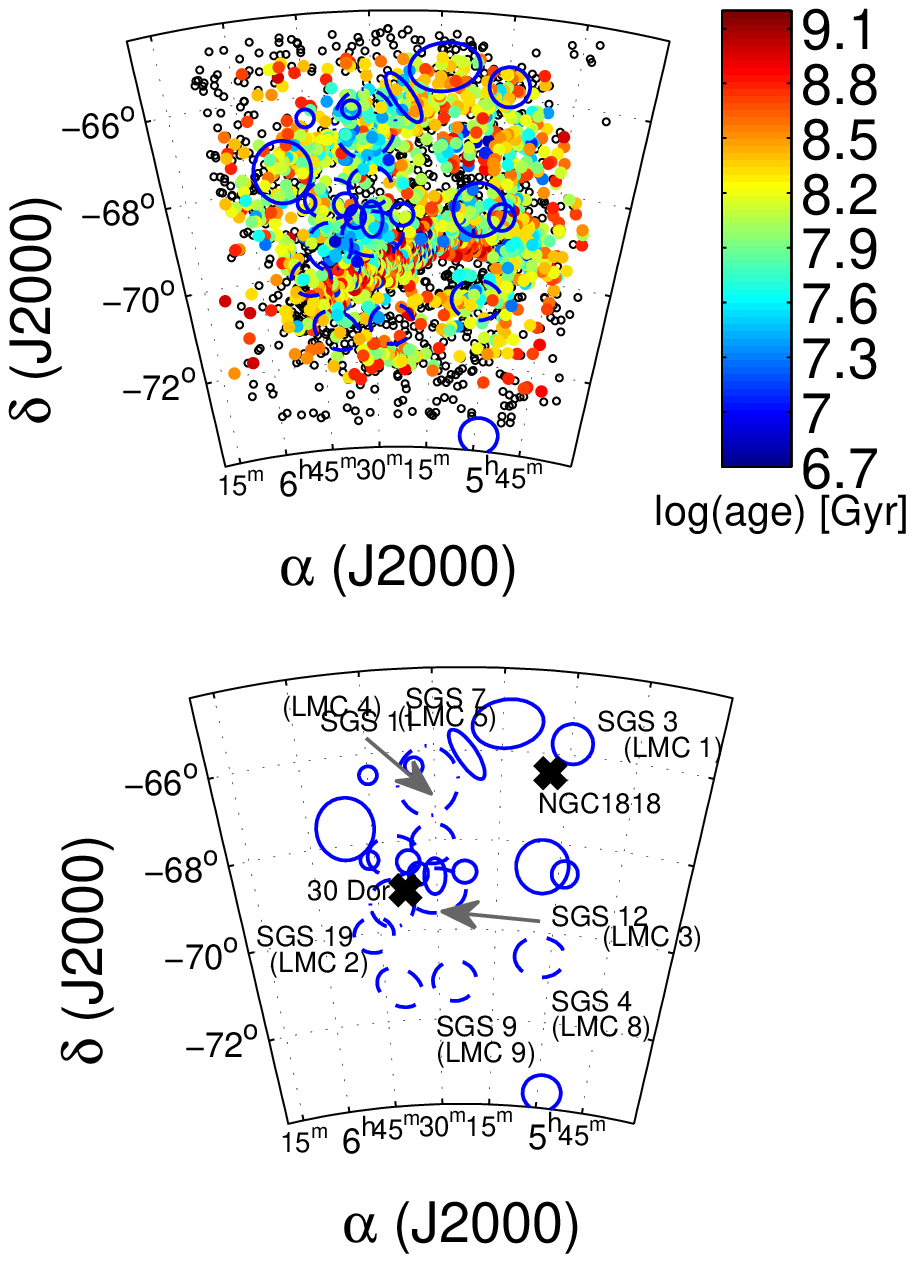}}  
 \caption{Spatial distribution of the combined sample of this work and PU00 of the young LMC star clusters. The blue 
 ellipses indicate the location of the 23 HI super-shells \citep[single SGSs (solid), complex SGSs (dashed), propagated 
 SGSs (dash-dotted)][]{kim99,kim05}. Marked are all shells discussed in the text, SGS 3 (LMC 1), SGS 4 (LMC 8), 
 SGS 7 (LMC 5), SGS 9 (LMC 9), SGS 11 (LMC 4), SGS 12 (LMC 3), and SGS 19 (LMC 2). The open circles indicate all objects 
 from B08 including a 'C'-classification (in total 4089 objects). The black crosses denote the locations of 
 30~Doradus and NGC~1818 (from left to right). North is up, west to the right.}
 \label{fig:lmcageall}
\end{figure*}

The age distribution of the SMC star clusters derived in our study shows a first peak of enhanced cluster formation around 160~Myr 
and a second one around 630~Myr. Only clusters with an age uncertainty smaller than class 3 were used. C06 found two main episodes 
of enhanced cluster formation between 5-15~Myr and at 90~Myr. Their cluster 
sample contains ages for 164 associations, which is the reason why their sample reaches down to ages of $\sim$3~Myr. 
In the combined sample three episodes of enhanced cluster formation are detectable, the youngest around 6.5~Myr, followed by a 
second peak around 160~Myr, and a third around 630~Myr. The analysis in the age range of the oldest peak is limited by the 
depth of the photometry, while the youngest peak mostly comes from the associations adopted from C06. 
The difference between the oldest peak found in our study and by C06 lies within the given uncertainties for the cluster ages 
($\sigma_{log(age)}$=0.3). 

For the LMC, PU00 detected three periods of enhanced cluster formation at about 7~Myr, 125~Myr, and 800~Myr, which are 
similar to the findings of G95. No associations and nebulae were considered in the study of PU00 and therefore the age distribution
of the combined LMC sample does not reach as young an age as the combined SMC sample. The youngest clusters in PU00's sample have 
ages of $\sim$5~Myr, but come with high age uncertainties as stated by the authors. The age determination of the oldest clusters 
in our sample with ages close to 1~Gyr is highly uncertain (class 3). Clusters with uncertainties smaller than class 3 were discarded 
and therefore the oldest peak is very weak. In the combined sample an increased number of clusters appears at $\sim$9~Myr and 
630~Myr, which is in very good agreement with G95 and P00. The most enhanced peak occurs at about 125~Myr. 

In both the SMC and the LMC, we find evidence for periods of enhanced cluster formation, which appear to have occurred during 
the same periods. The peaks at $\sim$125 and 160~Myr, respectively, are very pronounced in both galaxies and are probably 
correlated. The difference between the peaks is within the given uncertainties for the cluster ages (difference in log(age)=0.1). 
Model calculations performed by e.g., \citet{Bekki05} and \citet{Kalli06a,Kalli06b} showed that the MW, the LMC, and 
the SMC have only interacted long enough to produce the Magellanic Stream. According to these models, the last close 
encounter between SMC and LMC occurred about 100-200~Myr ago. The SMC star formation rate increases, if the LMC orbit leads to 
a close encounter with the SMC and vice versa. The star formation rate decreases again when the LMC recedes from SMC, thus leading 
to episodic cluster formation. Hence, this increase in cluster formation some 100-200~ Myr ago may have been triggered by a tidal 
interaction with the neighboring galaxy \citep[e.g., ][]{gardiner96,Bekki05,Kalli06a,Kalli06b}. The youngest peaks, 
however, might have another origin. Probably they are caused by associations, which did not yet dissolve. 
High velocity cloud-cloud collisions are another trigger mechanism of cluster formation 
\citep{zhang01,bekki04}. These collisions are particularly effective during galaxy interactions and mergers. High-speed motions 
may produce a high-pressure environment that in return can trigger turbulences or shocks \citep{elmergreen97}. Stellar winds and 
supernova explosions can also trigger star formation through compression by turbulent motions \citep{larson93}.

\subsection{Spatial Age Distribution}
\label{sec:youngies_spatial}

\subsubsection*{SMC:}

The spatial distribution of the star clusters in Fig.~\ref{fig:smcageall} clearly shows that the youngest 
clusters are located in the two large HI supershells 37~A and 304~A \citep{stanimirovic1999}. 
These features are believed to have their origin in the evolution of an OB association in which the most massive 
stars become supernovae and/or  where stellar winds lead to a runaway expansion and the formation of a supergiant shell 
\citep[e.g.,][]{bruhweiler80,tomisaka81,elmergreen82}. Most of the shells and super-shells are associated with young 
objects \citep{mccray87}. Inside the shells, second generation star clusters may form due to supernova explosions 
in the first generation clusters. 

Our (or the C06) data do not cover the western side of supershell 37~A, but most of the remaining objects are age-dated. 
We confirm the findings of C06 about the discontinuity in the spatial distribution of younger clusters in the supershells 
37~A and 304~A: That is, the youngest objects are found toward the eastern rim of the shell (37A) where gas 
and dust are located \citep{staveley97,stanimirovic1999} and toward regions luminous in H$\alpha$. Such HII regions 
visible in H$\alpha$ indicate the presence of massive stars (e.g., OB-stars, supergiants, luminous blue variable stars), 
supernova remnants, or diffuse ionized nebulae. The youngest objects of the shell 304~A are located toward its northern 
and western rim (toward 37~A) and also in the central part. The compression of gas due to shell-shell interactions 
related to the expansion of the shells might have triggered the cluster formation at the opposing rims as well as in 
the inter-shell region. H$\alpha$ is concentrated in the northern part of 304~A where we also find the youngest objects. 

In Fig.~\ref{fig:smcageslides} four snapshots are shown displaying the spatial distribution of star clusters within different
age ranges. In the first snapshot only clusters younger than 16~Myr are shown. The clusters are distributed along the 
SMC bar in an elongated but narrow area. The two supershells and the inter-shell region are clearly visible in this plot.
Clusters with ages between 16~Myr and 63~Myr are also concentrated along the bar, but there are outliers found toward the 
SMC wing or slightly offset of the bar (second snapshot). The supershell 304~A is highly populated by clusters of this
age range. Widely spread over the entire SMC main-body are clusters with ages between 63~Myr and 315~Myr (third snapshot). 
Moreover, the northern part of the SMC body is covered with clusters in this age range. Finally, the last snapshot displays 
clusters older than 315~Myr (up to $\sim$ 1~Gyr). These clusters mainly populate the western part of the SMC main body and 
only a few are found in the north or in the east. Supershell 304~A contains fewer of these older objects than 37~A and they 
are mainly located at the western rim of the shell, while in 37~A they are widely distributed. Generally, the cluster 
distribution in 304~A indicates a continuous cluster formation from a few Myr to 1~Gyr.

According to the standard model of shell formation \citep{mccray87} the older objects are supposed to be found in the 
center of the shell while younger objects are distributed around the edge. As mentioned above, the younger objects in 37~A
and 304~A are clustered toward the eastern and the western rim of the shells, respectively. Shell interaction due to 
collisions following shell expansion has probably triggered the cluster formation in the inter-shell region as well as 
at the two opposing rims. The reason might be that the shell is surrounded by an interstellar medium with differing densities. 
Shell expansion into higher-density regions leads to slower expansion; the corresponding strong 
gas compression leads to enhanced star formation in that region. On the other hand, if the shell expands into a region with 
lower density the star formation rate will be lower since there is less compression and less material to be swept up. 

We confirm C06's finding that the location of the young SMC clusters is correlated with HI intensities, which decrease with
increasing age. \citet{mizuno01a} found that young emission objects are positionally well correlated with CO clouds,
while emissionless objects of ages $\sim$6-100~Myr do not show any correlation to the CO clouds. Emission objects were 
excluded from our sample and therefore we cannot investigate their possible association with CO clouds.

\subsubsection*{LMC:}

The youngest star clusters are found mostly in the LMC HI supergiant shells (SGSs) 30 Doradus, SGS 7 (LMC 5), SGS 11 (LMC 4), 
SGS 12 (LMC 3), and SGS 19 (LMC 2) and in the 103 giant shells published by 
\citet{kim99,kim05} \citep[see also][]{davies76,meaburn80}. The authors classified the shells into three categories; 
the {\it single SGSs} are simple expanding stalled shells (solid), {\it complex SGSs} consist of smaller interlocked shells 
and their rims often contain smaller shells (SGS 4, 9, 10, 12, 17, 18, 20; dashed), and {\it propagated SGSs} are surrounded 
by another SGS (SGS 11, 19; dash-dotted). In most of the complex SGSs (see Fig.~\ref{fig:lmcageall})
star clusters are located along the rims and lack the youngest clusters. The smaller shells along their rims might indicate 
that triggered star formation occurred recently. The propagated SGSs may have formed by sequential star formation. 

De Boer et~al. (1998) proposed a scenario in which star formation in the LMC is triggered by the gas being compressed at the 
leading edge due to the bow-shocks as the LMC moves through the gaseous Galactic halo around the Milky Way. The bow-shock 
compresses large areas which leads to star formation on a large scale. Due to the LMC's rotation, the material at the leading 
edge will move away clockwise and distribute itself around the LMC's rim. In this scenario one would expect to find a progression 
in the age of star clusters along the direction of the rotation (in Fig.~\ref{fig:lmcageall} the galaxy rotates clockwise). The 
LMC disc shows solid-body rotation and has an approximate full rotation velocity for a small inclination of $\sim$150~kms$^{-1}$ 
at 1.5~kpc \citep{deBoer98}. The time of one full rotation period is $\sim$250~Myr. If present the progression of age along 
the galaxy's rim should be detectable. The youngest shell structures lie in the proximity of 30 Doradus, the largest star 
forming region of the LMC. 30 Doradus lies close to the leading edge of the LMC. It is the largest star forming region in the 
galaxy and may very well be related to the bow shock. 

%\begin{figure}
% \resizebox{\hsize}{!}{\includegraphics{PA.eps}}  
% \caption{Cluster age distribution inside the MCs. Cluster age distribution in a 360$^{\circ}$ angle around the center coordinates
% of the SMC and LMC adopted by \citet{karachentsev04}.
% We find no evidence for a progression in star cluster age in the direction of the rotation of the LMC as proposed by 
% \citet{deBoer98}. Because of the absence of rotation of the SMC no such age distribution is expected.}
% \label{fig:PA}
%\end{figure}

\citet{grebel98} used 
Cepheids and other supergiant stars to study the recent star formation history of the LMC. They found that the majority 
of objects younger than 30~Myr are concentrated on the south-eastern border, while others are widely distributed across 
the entire disk which cannot be explained with the bow-shock star formation model. A progression of age was found in several 
giant shells along the LMC rim moving from the south-east (LMC~2) to the north (NGC~1818 at $\alpha$ = 
$5^h04^m03^s$, $\delta$ = $-66^{\circ}26'00''$) \citep{deBoer98}. 30 Doradus has an age of 3-5~Myr, LMC~4 of 9-16~Myr, and 
NGC~1818 of 25-30~Myr \citep{Parker93,Will95,Braun97,Grebel97}. The difference in age between these shells corresponds 
to their distance along the border of the disk divided by the galaxy's rotation velocity \citep{mastropietro09}. 

We find no evidence for such a cluster age distribution for the LMC. The older clusters are 
mostly located in the bar region and a few are located along the rim. Most clusters populating the LMC bar were adopted 
from the sample published by PU00. Ages of clusters older than $\sim$1~Gyr cannot be derived due to the limited 
photometric depth of the MCPSs, which does not resolve MSTOs of older clusters. Overall, our spatial cluster age distribution 
does not support ram pressure as the primary agent of cluster formation in the LMC. 

\begin{figure*}
 \resizebox{\hsize}{!}{\includegraphics{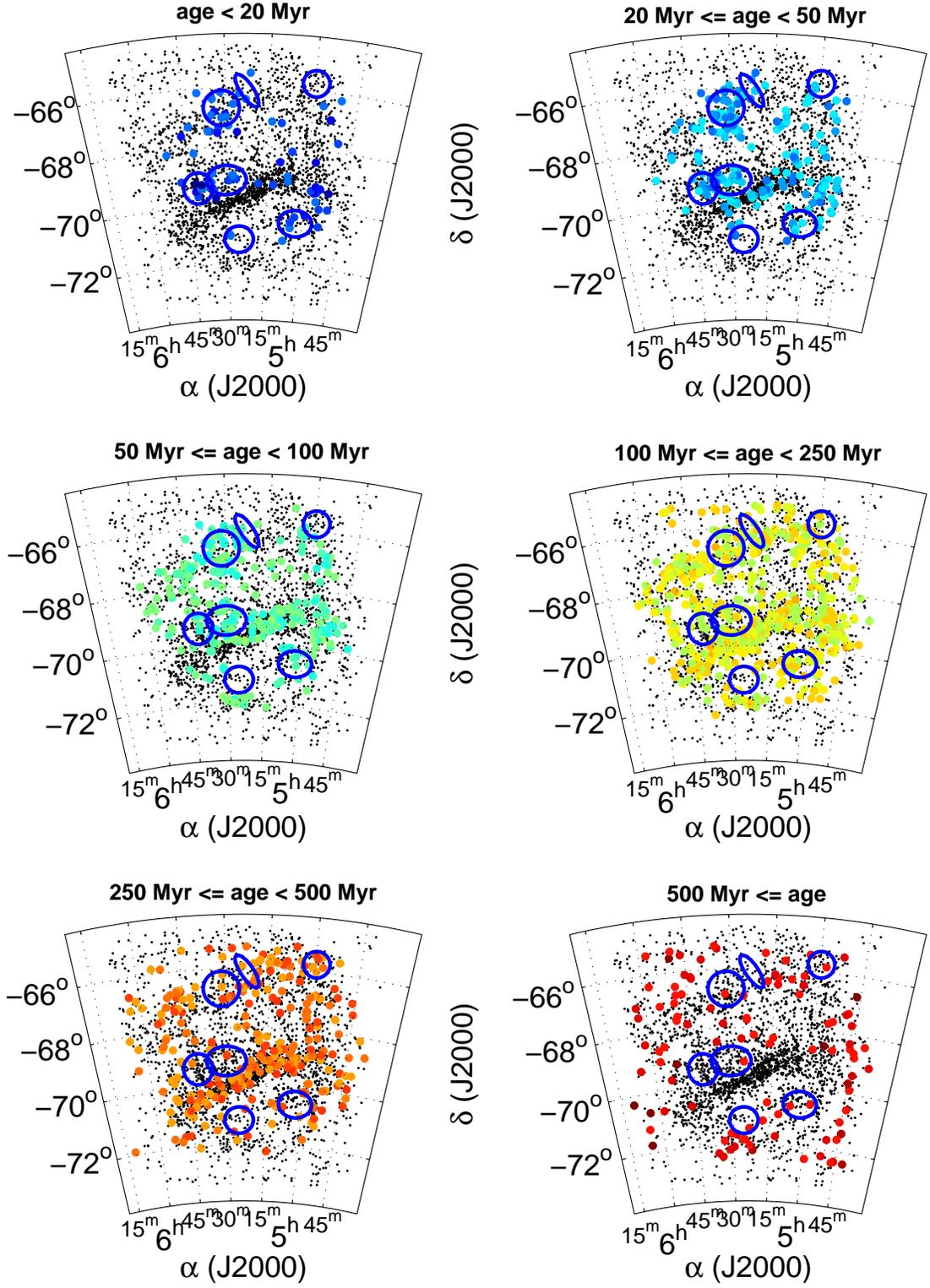}}  
 \caption{Spatial distribution of LMC clusters younger than 1~Gyr in different age bins. Note how star formation 
 progresses along the bar and how previously inactive regions like 30~Dor gradually become centers of cluster formation
 activity.}
 \label{fig:lmcageslides}
\end{figure*}

As mentioned above, the highest concentration of young clusters can be found in the 30 Doradus region (NGC\,2070) in 
the southeastern part of the LMC at roughly $\alpha$ = $5^h38^m$, $\delta$ = $-69^{\circ}06'$. In this large star 
forming complex many SGSs are interlocking, where active star formation occurred simultaneously \citep{kim99}. A number 
of smaller shells have formed along the rims of the supershells, as expected for self-propagating star formation 
\citep[e.g., ][]{mccray87}. The HII regions are sometimes associated with the HI shells and the largest concentration can 
be found in the 30 Doradus region as well as in SGSs hosting the youngest star clusters. 

Fig.~\ref{fig:lmcageslides} displays six snapshots showing the spatial distribution of LMC star clusters within different 
age bins. The first snapshot shows only clusters younger than 20~Myr. The star clusters are located in the 30 Doradus 
region, LMC~4, in the western part of the bar, and along the rim in the southwestern part of the galaxy. 
Star clusters within the age range of 20~Myr and 100~Myr in the second and third snapshot are more widely distributed 
over the entire SGSs and in their inter-shell regions as well as in the giant shells. There is no one-to-one correlation 
between the HI shells and the ionized gas traced out by the H$\alpha$ regions \citep{mizuno01a, book08}. Along the bar, 
we now see pronounced activity in the western region and in the center, while the eastern part of the bar is essentially 
quiescent. Note also the particularly strong activity in the 30~Doradus region and in and around SGS~11 (LMC~4). 

The LMC bar and the LMC rim are fully populated with star clusters within the age range of 100~Myr and 250~Myr (fourth 
snapshot). Star clusters with ages between 250~Myr and 500~Myr mostly populate the LMC bar, but many are also distributed 
along the western and northwestern LMC rim as well as in the northeast (fifth snapshot). Star clusters older than
500~Myr (up to 1~Gyr) are mostly distributed along the LMC rim, while the bar region is essentially devoid of clusters 
in this age range. Also for the LMC, we find the older objects outside the shell regions and the younger objects are 
located along the rims and inter-cloud regions. The formation of young LMC clusters may have been triggered by shell 
expansion and interactions, which might be the reason why we do not find evidence for the standard model of shell 
formation (McCray \& Kafatos 1987, see also Braun et al. 1997, Points et al. 1999). In the following we will discuss 
the spatial distribution of the clusters of different ages inside the supershells SGS~3 (LMC~1), SGS~4 (LMC~8), SGS~9 
(LMC~9), SGS~11 (LMC~4), SGS~7 (LMC~5), SGS~12 (LMC~3), and SGS~19 (LMC~2) in more detail. 

\textit{SGS~3} (LMC~1) is located in the northwestern corner of the LMC at $\alpha$ = $5^h00^m$, $\delta$ = 
$-65^{\circ}40'$ and has a dynamical age of $\sim$7~Myr \citep{kim99}. Blue and red supergiants in the southern shell 
area were found to have ages less than 30~Myr \citep{grebel98}. Young stellar objects are mostly distributed around 
the shell's southern periphery \citep{book09} inside nearby HII regions. There is no cluster younger than 30~Myr in our 
sample in SGS~3 (LMC~1) and only a few clusters within the age range of 30~Myr-100~Myr are located in the southeastern 
region. Clusters with ages between 30~Myr and 500~Myr are distributed over the entire projected area of shell. The 
clusters from our sample in this shell have a mean age of $\sim$150~Myr. Star formation in this shell
mostly occurred in HII regions independent of the shell \citep{book09}. One OB association stretches from the center of 
the shell to a bright southeastern HII region indicating a propagation of star formation. No indication of self-propagating 
star formation can be found in this study, which might be due to incompleteness of the sample. Given this age structure, 
the formation of the shell is unrelated to the comparatively much older star clusters. 

\textit{SGS~4} (LMC~8) lies in the southwestern corner of the LMC at $\alpha$ = $5^h03^m$, $\delta$ = $-70^{\circ}30'$ 
and has a dynamical age of $\sim$6~Myr \citep{kim99}. This shell is clearly visible in the spatial distribution of 
supergiants (ages $<$ 30~Myr, Grebel \& Brandner 1998). OB associations were found in the north and the east of the 
shell. SGS~4 (LMC~8) has a complex structure of HI lobes, which are connected \citep{book08}. Star clusters in this 
shell cover an age range between $\sim$12 to 500~Myr and have a mean age of 100~Myr. The northern, the north-eastern, 
and the eastern lobes show expansion where also a few clusters younger than 50~Myr are located (see 
Fig.~\ref{fig:lmcageslides}). The southern lobes show no distinct expansion \citep{book08} and in this region the shell 
appears to be essentially quiescent. A few clusters in the age range between 50~Myr and 100~Myr are found in the center 
and the norther region of the shell. Clusters in the age range betweeen 100~Myr and 250~Myr are located in the center, 
the east, and the northwest. The shell is mostly devoid of clusters older than 250~Myr. Only a small number of clusters 
is located along the rim regions in the east and the west. As for SGS~3 (LMC~1), no indication for self-propagating 
star formation is found in this study.

\textit{SGS~9} (LMC~9) is located in the southern part of the LMC at $\alpha$ = $5^h25^m$, $\delta$ = $-70^{\circ}05'$. 
The shell is a collection of several giant H$\alpha$ shells, but has no corresponding HI shell structure, which is why 
\citet{book08} characterized the shell as \textit{false}. This shell has almost never shown cluster formation activity 
within the period studied here. HI supershells are distributed from west to northeast along the periphery. The shell  
showing the most cluster formation activity is a region in the southeast, outside the boundary of the shell. In this region many 
clusters with ages between 50~Myr and 250~Myr are found, which might be correlated with the expanding H$\alpha$ and HI 
region. The few clusters younger than 50~Myr are located at the northeastern shell rim, which might be associated with 
HII regions. This shell is probably very old, because almost no young clusters are located in this shell and barely any 
supergiants and Cepheids were found in the shell region \citep{grebel98}. The star clusters from our sample in this shell 
have ages between $\sim$20~Myr to 1~Gyr with a mean age of 170~Myr.

\textit{SGS~11} (LMC~4) is the largest SGS in the LMC and is located at $\alpha$ = $5^h32^m$, $\delta$ = $-66^{\circ}40'$. 
The shell is in collision with SGS~7 (LMC~5). The densest part of the shell lies at the northwestern border in the region 
between SGS~11 (LMC~4) and SGS~7 (LMC~5). Multiple OB associations exist in HII regions in the interaction zone, along the 
shell rim, and near the center of the shell \citep{book09}. \citet{yamaguchi01b} found the clusters at the rim of SGS~11 
(LMC~4) to be associated with CO clouds. They are mostly noticable in the interaction region. Stars inside the shell were 
found to be $\sim$9-16~Myr old, while those at the rim are $\lesssim$6~Myr old \citep{Braun97}. The analysis of Cepheids 
and other supergiants showed that star formation in this shell started about 25~Myr ago in its center region and then slowly 
moved out toward the outer rims \citep{grebel98}. SGS~11 (LMC~4) appears to be the only shell that shows a weak indication 
of propagating star formation and has formed only within the last 10~Myr. Star clusters from our sample cover an age range 
between $\sim$12 and 500~Myr. The mean age lies at 60~Myr. Star clusters with ages less than 20~Myr are found along the 
shell rim where also the HII regions were identified \citep{book09}. Star clusters with ages between $\sim$50-250~Myr are 
widely distributed across the entire shell. Clusters older than 250~Myr are mostly found at the southern rim of the shell. 
Two are found in the center region and a single one at the northern rim. If the very young age of this shell is correct, the 
formation of clusters older than $\sim$20~Myr is not related to SGS~11 (LMC~4). No indication for self-propagating star 
formation is found in this study.

\textit{SGS~7} (LMC~5) is located to the northwest of SGS~11 (LMC~4) at $\alpha$ = $5^h22^m$, $\delta$ = $-66^{\circ}00'$ 
and has a dynamical age of $\sim$5~Myr \citep{kim99}. The shell is in collision with SGS~11 (LMC~4), which caused a density 
enhancement on the eastern side of the shell. The cluster age distribution is similar to the one in SGS~11 (LMC~4). The 
star clusters from our sample in this shell have ages between 20 to 400~Myr with a mean age of 130~Myr. Only five 
clusters from our sample younger than 100~Myr are found in this shell and three of them are located at the eastern rim. 
No OB associations are found inside the shell, but along the interaction zone and along the rim \citep{book09}. No OB 
association was detected in the shell interior. The shell has a relatively large CO concentration \citep{yamaguchi01a}. 
The highest CO density is located in the interaction zone, superposed on OB associations and corresponding HII regions. 
Two more high-density CO regions are located in the southern region of the interaction zone and in the north. Only very few
objects younger than 20~Myr were found in the present sample because of our object selection. The southern part of 
the shell and the surroundings are covered by star clusters with ages between 100-500~Myr. The shell is devoid of objects 
older than 500~Myr. Like SGS~11 (LMC~4), this shell probably is very young. Especially in the interaction zone, where SGS~7 
(LMC~5) and SGS~11 (LMC~4) are colliding, a large quantity of blue and red supergiants are located indicating recent star 
formation \citep{grebel98}. The old clusters in the south of the shell might have formed in a shell that is dissolved 
already or via a non-shell related mechanism. Star formation in this shell is probably triggered by shell expansion, 
especially in the collision zone.

\textit{SGS~12} (LMC~3) is located at $\alpha$ = $5^h30^m$, $\delta$ = $-69^{\circ}00'$ and is in collision with SGS~19 
(LMC~2). The large star forming region 30 Doradus lies in the interaction zone. The SGS is divided into northern and 
southern lobes and shows significant HI and H$\alpha$ structures \citep{book08}. Old OB associations (age$\gtrsim$12~Myr) 
exist in the interior of the shell, while young OB associations (age$\lesssim$12~Myr) are found within dense HII regions. 
The cluster mean age of this shell lies at $\sim$75~Myr and they cover an age range of 5~Myr to 1~Gyr. Clusters younger 
than 20~Myr are concentrating in the center region of the shell. Clusters with ages between 20-250~Myr are more widely 
distributed. The shell barely contains any clusters older than 250~Myr. No signs of propagating star formation events are 
found. Clusters might have formed via shell expansion, most efficiently in the collision zone. Looking at the age 
distribution of blue and red supergiants (age$<$30~Myr) the intershell region is clearly visible as a region with a high
concentration of these objects \citep{grebel98}. The oldest supergiants may have triggered the subsequent compression of 
dense molecular clouds and resulting star bursts. 

\textit{SGS~19} (LMC~2) is located to the southeast of 30 Doradus at $\alpha$ = $5^h44^m$, $\delta$ = $-69^{\circ}20'$
and is the brightest SGS in the LMC in H$\alpha$ \citep{kim99}. The shell is colliding with SGS~12 (LMC~3). The shell 
has very complex structures in HI and H$\alpha$ and shows star forming regions on its western edge \citep{book08}. The 
clusters from our sample cover an age range between 40~Myr and 1~Gyr. The cluster mean age lies at $\sim$250~Myr. Clusters 
younger than 50~Myr are located in the center of the shell and toward 30 Doradus. The shell hosts only a few clusters with 
ages between 50~Myr and 250~Myr and is almost devoid of older clusters. 29 clusters in the direction of LMC~2 were reported 
for which ages have been determined \citep[e.g., ][]{els85,bica96}. Eleven of them are younger than 10~Myr, seven have
ages between 10-30~Myr, one has an age of 30-70~Myr, and one is significantly older with an age of 400-800~Myr. The 
youngest clusters are associated with HII regions that are located along the periphery of LMC~2 \citep{Points99}. The 
analysis of supergiants revealed that this shell is very young (10-15~Myr, Grebel \& Brandner 1998) and in the spatial 
distribution of supergiants the interaction region between SGS~19 (LMC~2) and SGS~12 (LMC~3) is clearly visible.

\citet{mizuno01b} found that most of the CO clouds in the LMC are distributed in dense parts of HI gas and over several 
HII regions. LMC CO clouds are often located at the outer edge of SGSs as for e.g., SGS 11 (LMC~4) \citep{yamaguchi01a}.  
Furthermore these authors  found that clusters associated with the CO clouds are younger than 5~Myr, which we cannot 
investigate due to the limited age range of the sample.

\begin{figure}[h!]
 \resizebox{\hsize}{!}{\includegraphics{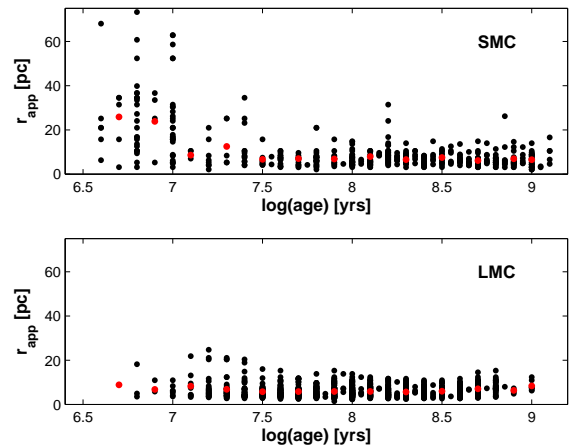}}  
 \caption{Age vs apparent radius. In both SMC and LMC, no indication of cluster dissolution is 
 visible. The large radii for associations in the r vs age distribution of the SMC come from C06. The red dots represent the 
 mean of the radius distribution.}
 \label{fig:dissolution}
\end{figure}

\subsection{Dissolution Effects}
\label{youngies_dissolution}

Due to the different morphology of the MCs compared to the MW and the absence of differential rotation 
of SMC and LMC, low-concentration clusters can easily survive in these two galaxies, while the MW rotates differentially, 
which leads to shear and interaction with massive molecular clouds that can result in cluster dissolution. For the LMC 
\citet{parmentier08a} found a cluster formation rate of 0.3 clusters~Myr$^{-1}$ 5~Gyr ago, which increased steadily to 
a present rate of $\sim$25 clusters~Myr$^{-1}$. For the SMC such studies have not been published yet. The main distinct 
phases and corresponding typical timescales of cluster disruption in the MCs are: (I.) {\it infant mortality} 
($\sim$10$^7$~yr), (II.) mass loss dominated by {\it stellar evolution} ($\sim$10$^8$~yr), and (III.) a phase dominated 
by {\it tidal relaxation} ($\sim$10$^9$~yr) in which the mass loss is driven by the clusters' dynamical evolution and 
external influence of the tidal field of the host galaxy \citep[][and references therein]{lamers05,Bastian08}. 
Additionally, tidal external perturbations speed up the process of disruption during all three phases, but these effects 
operate on longer timescales and mostly affect phase III. Most of the clusters in our sample have survived the first phase 
and are dominated by the second and third phase. The rate of infant mortality is highly dependent on the ambience of the 
host galaxy, but is largely mass-independent - at least for masses in excess of $\sim10^4~M_\odot$ \citep[e.g., ][]{Bastian05}. 
Infant mortality is mainly driven by gas expulsion. This lifecycle of a cluster results from the expulsion of the 
intra cluster gas due to explosive expansion driven by stellar winds or supernova activity 
\citep[e.g., ][]{Mengel05,Bastian06b,goodwin06}. The end of cluster infant mortality depends on the crossing time 
of the cluster's gaseous progenitor and on the external tidal field \citep{Parmentier09}. The denser the gaseous precursor
of the star cluster, the shorter the cluster's crossing time and the quicker the cluster's violent relaxation. The 
stronger the external tidal field, the smaller the cluster's tidal radius and the faster unbound stars leave the cluster. 
Because the external tidal fields of the MCs are not very strong, most star clusters in the MCs surviving the phase of 
infant mortality will have returned to an equilibrium state after $\sim$40-50~Myr \citep{goodwin06}. Therefore, some 
of the clusters in the present sample might dissolve and not survive the first phase of cluster dissruption, while most 
clusters have ages of 40~Myr and higher, and therefore are in the phase where stellar evolution has become a major agent 
of change, whereas the gas has already been expelled. 

The second (mass-dependent) lifecycle of a cluster includes the so-called secular evolution (gas free evolution). This 
phase is driven by 2-body relaxation and the morphology of the host galaxy. Therefore, the disruption time of a cluster 
is dependent on the internal cluster conditions, such as the initial mass, density, and velocity dispersion, as well as 
on external conditions, such as the orbit in the galaxy and tidal heating 
\citep[e.g., ][]{boutloukos03,lamers05,parmentier08a}. 

%While the disruption timescale in a spiral 
%galaxy like the MW is rather short and about 90-95\% of all star clusters get disrupted, most star clusters in an irregular 
%galaxy can survive (e.g., 70\% of all SMC clusters and 90\% of all LMC clusters, de Grijs \& Goodwin 2008/2009). \\
Fig.~\ref{fig:dissolution} shows the age-radius relation of the combined SMC and LMC cluster sample with ages between 
$\sim$5~Myr and 1~Gyr. SMC and LMC cluster ages are plotted against their apparent overall radii (Eq.~1) from the 
combined samples derived in this 
study and in C06 (SMC) and PU00 (LMC). The youngest objects in the SMC have radii up to $\sim$4~arcmin, but these objects 
are clusters adopted from C06 and are mostly classified as associations. Moreover, until an age of $\sim$40~Myr star 
clusters in the MCs can still undergo infant mortality \citep{goodwin06} and therefore we have to treat this subsample 
with caution. The radii decrease until an age of $\sim$25~Myr and then remain relatively constant until 1~Gyr. Because the 
youngest objects in the LMC were not age-dated no extended objects can be seen at the young age end of the distribution. 
The cluster radii in the LMC show the same behavior as the SMC clusters and are constant with time during the analyzed 
period. The cluster disruption time for the SMC and the LMC derived in the literature is of the order of $8 \times 10^9$~Myr 
\citep[e.g.,][]{boutloukos03,parmentier08a}. 

We have shown \citep{glatt09} that clusters with ages older than 1~Gyr have a trend for larger core radii with increasing 
age. This trend was first found by \citet{mackey03b} and is also visible for LMC, Fornax, and Sagittarius clusters 
\citep{mackey03a,mackey03c}. We note, however, that the radii in Fig.~\ref{fig:dissolution} are \textit{apparent overall}
radii (see Section~\ref{sec:youngies_ages}) so that we cannot draw any firm conclusions on possible dissolution effects based on the 
observed sizes alone.

\section{The cluster luminosities}
\label{sec:youngies_lum}
\vspace*{0.3cm}

\begin{figure}
 \resizebox{\hsize}{!}{\includegraphics{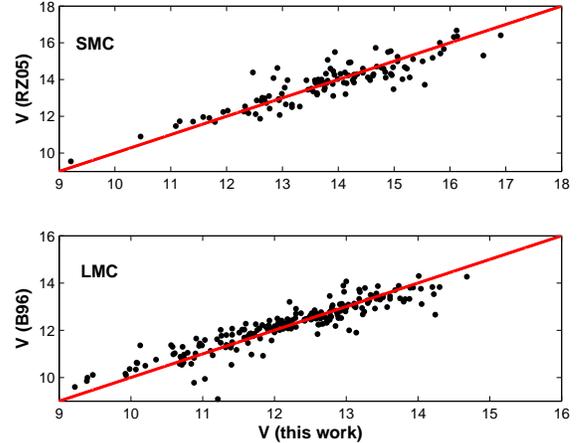}}  
 \caption{V band magnitudes derived in this work are compared with the V band magnitudes derived by RZ05 (SMC) and by 
 B96 (LMC) for the clusters in common. The solid line is unity.}
 \label{fig:Vrafmevergl}
\end{figure}

RZ05 published luminosities of 204 SMC star clusters measuring integrated colors from the MCPS. B96 published 
integrated photometry of 624 LMC star clusters that was based on observations carried out at the CTIO in Chile 
and at the CASLEO in Argentina. For the age-dated clusters in our samples, we computed luminosities by summing 
the flux of each star within the individual cluster radii (Eq.~1). To correct for field stars, the 
luminosity of a concentric annulus between 2 and 1~arcmin was calculated for each cluster, area corrected, and 
subtracted from the cluster luminosity. 112 clusters of our sample are in common with RZ05's sample and 217 
clusters are in common with B96's sample. The comparison between the V band luminosity derived in our study with 
RZ05 and B96 (Fig.~\ref{fig:Vrafmevergl}) shows that the V band magnitudes are overall in good agreement. 
The LMC V band magnitudes $<$12~mag published by B96 are systematically fainter by a small amount than the ones computed 
in this study. On the other hand, at the faint end the V band magnitudes (V$>$13.50~mag) by B96 appear to be 
slightly brighter. The dispersion about the 1:1 relation (red solid lines) for all clusters in common is 
$\sigma_V$ = 0.28 (SMC) and 0.22 (LMC).

\begin{figure}[h!]
 \resizebox{\hsize}{!}{\includegraphics{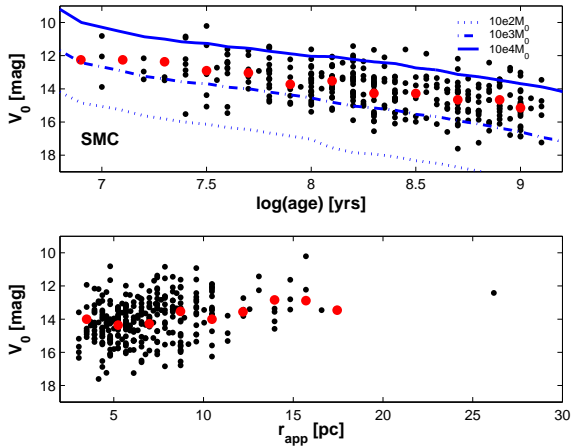}}  
 \caption{$V_0$ vs log(age) and $V_0$ vs apparent overall radius for our SMC clusters. The luminosities were derived in 
 this study. The red dots represent the mean of the age distribution in magnitude bins of 0.3~mag. The trend for 
 fainter magnitudes with increasing age is obvious. The blue lines represent three GALEV models \citep{Kotulla09}
 of different total mass: $10^4M_{\odot}$ (solid), $10^3M_{\odot}$ (dash-dotted), and $10^2M_{\odot}$ (dotted).}
\label{fig:smcraf_Vager}
\end{figure}

\begin{figure}[h!]
 \resizebox{\hsize}{!}{\includegraphics{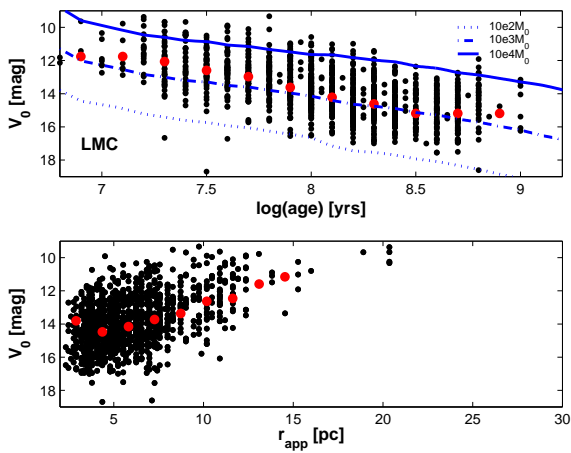}}  
 \caption{$V_0$ vs log(age) and $V_0$ vs apparent overall radius relations for our LMC clusters. The luminosities 
 were derived in this study. The red dots represent the mean of the age distribution in magnitude bins of 0.3~mag. The trend 
 for fainter magnitudes with increasing age is also obvious in this plot. The blue lines represent three GALEV models 
 \citep{Kotulla09} of different total mass: $10^4M_{\odot}$ (solid), $10^3M_{\odot}$ (dash-dotted), and $10^2M_{\odot}$ (dotted).}
 \label{fig:lmcraf_Vager}
\end{figure}

The younger the star clusters the brighter they are because young clusters still contain very massive hot stars (e.g, 
supergiants), which contribute most of the light. The older a cluster gets the more of these massive stars are in 
their end stage of evolution and no longer contribute to the cluster light. Therefore, clusters become fainter and redder with 
increasing age. This effect can be seen in the upper panels of Figs.~\ref{fig:smcraf_Vager} (SMC) and~\ref{fig:lmcraf_Vager} 
(LMC), in which cluster age is plotted versus luminosity. The red dots represent the mean of the age distribution in magnitude 
bins of 0.3~mag. We overplotted three GALEV models \citep{Kotulla09} for the total cluster masses of $10^2M_{\odot}$ (blue 
dash-dotted line), $10^3M_{\odot}$ (blue dotted line), and $10^4M_{\odot}$ (blue solid line) using a Salpeter initial mass function. 
GALEV is a model for computing the spectral evolution of single stellar populations and galaxies. The GALEV models give 
an overview of the cluster mass range. The lower plots of 
Figs.~\ref{fig:smcraf_Vager} and~\ref{fig:lmcraf_Vager} show that star clusters of the SMC and LMC become brighter with 
increasing radius, which is expected assuming a larger number of stars within the cluster radius, contributing to the 
total cluster luminosity.

\section{Summary}
\label{sec:youngies_summary}

We have presented ages and luminosities of 324 and 1193 populous SMC and LMC star clusters, respectively. An age range 
of $\sim$9~Myr to 1~Gyr was covered based on isochrone fitting to resolved color-magnitude diagrams in both galaxies. 
Using only cluster ages derived in this 
study, we find two maxima of enhanced cluster formation for both galaxies which appear to be correlated. In the SMC, the 
peaks are found at $\sim$160~Myr and $\sim$630~Myr, and in the LMC at $\sim$125~Myr and $\sim$800~Myr. Model calculations 
predict that the last close encounter between LMC and SMC occurred around 100-200~Myr ago. During a close encounter, the star 
formation is expected to be enhanced. Therefore, the first peaks in the cluster age distributions 
could have been triggered by this tidal interaction. Extending our samples with cluster ages derived by C06 we find a third
pronounced period of enhanced cluster formation in the SMC at around 8~Myr. We find the same in the LMC combining our sample 
with the one of PU00. These peaks are only visible if we extend our sample with objects classified as associations, objects
which did not or could not reach higher ages because they dissolve too quickly.  

The youngest objects in both galaxies are associated with super giant shells, giant shells, the inter-shell 
region, and with HII regions. Their formation is probably related to shell expansion and shell 
interaction. In the spatial distribution of the clusters younger than $\sim$16~Myr the two SMC shells are clearly visible. The 
older objects are widely spread across the entire SMC main body, but show a concentration in the western part of the galaxy. 
In the LMC, the youngest objects are concentrated in 30 Doradus, SGS 11 (LMC~4), and in the giant shells located in the western
part and in the bar region. The older LMC clusters are mostly distributed along the bar and along the rim. One can see nicely
how star cluster formation propagated along the LMC bar. We find no indication
for propagating star cluster formation in the SGSs in either LMC or SMC. Most of the LMC star clusters are older than the
dynamical ages of the SGSs and therefore may have formed in shells, which already have dissolved and cannot be detected at 
the present day.

No obvious dissolution effects were found for MCs star clusters younger than $\sim$1~Gyr. It is quite difficult to ascertain 
a real absence of cluster dissolution using this study. Two biases may play a major role: 1. Infant mortality cannot be accounted 
for, because very young star clusters and OB-associations are not included in our sample; and 2. Cluster dissolution processes 
for clusters older than $\sim$1~Gyr, because we did not age-date clusters in this age range. Within the time period considered 
here - 10~Myr to 1~Gyr - we do not find evidence for cluster dissolution.

In both galaxies, the clusters become fainter with increasing age. The very massive hot stars, which are still present in the
young star clusters and contribute most of the light, become fainter and redder with increasing age and so do the star
clusters. This trend can be seen in both the LMC and the SMC. The total cluster luminosity increases with increasing radius
due to a larger number of stars within the cluster radius.

\begin{acknowledgements}
We thank the anonymous referee for extremely useful suggestions to improve our paper.
We gratefully acknowledge support by the Swiss National Science Foundation through grant number 200020-105260 and 
200020-113697. Andreas Koch acknowledges support by an STFC postdoctoral fellowship.
\end{acknowledgements}

{}
\end{document}